\documentclass[aps,pre,twocolumn,footinbib,longbibliography]{revtex4-2}

\usepackage{graphicx}
\usepackage{subfigure}
\usepackage{dcolumn}
\usepackage{bm}
\usepackage{times}
\usepackage{amsmath}
\usepackage{amssymb}
\usepackage{color}
\usepackage{multirow}
\usepackage[normalem]{ulem}

\begin{document}
\title{Modulating radiative heat and momentum transfer via the thermal Purcell effect}

\author{Liyao Jiao}
\affiliation{Graduate School of China Academy of Engineering Physics, Beijing 100193,
China}
\author{Yaohua Liu}
\affiliation{Graduate School of China Academy of Engineering Physics, Beijing 100193,
China}
\author{Gaomin Tang}
\email{gmtang@gscaep.ac.cn}
\affiliation{Graduate School of China Academy of Engineering Physics, Beijing 100193,
China}

\bigskip

\begin{abstract}
  The thermal Purcell effect describes the modification of the local density of states of
  the fluctuating electromagnetic field induced by a Fabry-P\'{e}rot cavity, leading to
  the enhancement or suppression of radiative transport quantities.
  Using fluctuational electrodynamics, we investigate nonequilibrium radiative heat,
  linear-momentum, and angular-momentum exchange between a magneto-optic nanoparticle and
  a Fabry-P\'{e}rot cavity. Analytical expressions for the spectral densities reveal that
  geometric confinement modifies the electromagnetic local density of states, producing
  distinct behaviors for different transport quantities. Specifically, sub-wavelength
  confinement enhances radiative heat and angular-momentum transfer, but suppresses the
  lateral force. Additionally, interference between cavity modes causes all transfer
  quantities to oscillate spatially with particle position. At the cavity center, mirror
  symmetry enforces a parity decomposition of electromagnetic fluctuations resulting in a
  vanishing lateral force, whereas heat transfer and torque remain finite through combined
  even and odd modal contributions. These results demonstrate that cavity engineering
  provides selective control over nanoscale energy and momentum transfer via structured
  electromagnetic fluctuations.
\end{abstract}

\maketitle

\section{Introduction}
Thermal radiation, a ubiquitous physical phenomenon traditionally described within the
framework of fluctuational electrodynamics, has long been investigated primarily for
energy exchange between objects at different temperatures~\cite{Pendry99, review05,
review07, review15, review18, review21}. Recent studies have expanded this scope to
demonstrate that thermal fluctuations can also mediate the transfer of mechanical
quantities, specifically linear and angular momentum~\cite{lateral_17, lateral_21,
lateral_21-2, lateral_23, lateral_23-2, GT24-2, lateral_25, GT26-1, GT26-2}. Crucial to
induce momentum transfer is the introduction of nonreciprocal effects, which break
time-reversal symmetry and bypass the constraints imposed by conventional thermodynamic
reciprocity. Nonreciprocity can be effectively achieved in various scenarios, most notably
by applying an external magnetic field to magneto-optic materials or by using magnetic
Weyl semimetals that possess intrinsic time-reversal symmetry breaking~\cite{WSM_radiate1,
WSM_radiate2, WSM_radiate3, WSM_radiate4, GT_WSM}.

While radiative heat and momentum transfer can be tailored by actively modifying the
intrinsic material properties, an alternative approach is to engineer the surrounding
electromagnetic environment itself via optical cavities. Analogously to the conventional
Purcell effect in quantum optics, where the spontaneous emission rate of an emitter is
modified by a resonant cavity, the thermal Purcell effect offers a route to modulate
radiative transport out of thermodynamic equilibrium within a cavity~\cite{cavity-23,
cavity-24, cavity-25, cavity-25-2}. Specifically, embedding a material system inside a
Fabry-P\'{e}rot cavity alters the local density of states (LDOS) of the fluctuating
electromagnetic field, thereby enhancing or suppressing specific radiative transitions.
Crucially, while a cavity-induced shift of electronic energy levels typically relies on
the strong light-matter coupling, the modulation of radiative heat flow operates even
under the weak light-matter coupling condition. 

In this work, we investigate the impacts of the cavity thermal Purcell effect on the
transfer of radiative heat, linear momentum, and angular momentum. Specifically, we
consider a system consisting of a spherical magneto-optic indium antimonide (InSb)
nanoparticle placed within a Fabry-P\'{e}rot cavity formed by two parallel perfectly
metallic plates under an external magnetic field [see Fig.~\ref{fig1}(a)]. By using the
fluctuational electrodynamics together with the cavity-modified dyadic Green's functions,
we derive expressions for the spectral densities of the radiated power, the lateral force,
and the torque.
We demonstrate the influence of the cavity length and particle-substrate separation 
on these transport spectral densities. Our results demonstrate that geometric 
confinement strongly enhances radiative heat transfer and radiative torque, while 
generally suppressing the nonreciprocal lateral force. Furthermore, these quantities 
exhibit pronounced spatial oscillations arising from interference between cavity 
modes.

The rest of this paper is organized as follows. In Sec.~\ref{sec:II}, we introduce the
photonic Green's functions and obtain the expressions for the electric LDOS within the
parallel-plate cavity. Section~\ref{sec:III} presents detailed analytical and numerical
results along with a discussion regarding the cavity-modulated radiative heat flow,
lateral force, and torque acting on the InSb nanoparticle. Our work is summarized in
Sec.~\ref{summary}.

\section{Photonic Green's function and local density of states} \label{sec:II}
\begin{figure}
\centering
\includegraphics[width=\columnwidth]{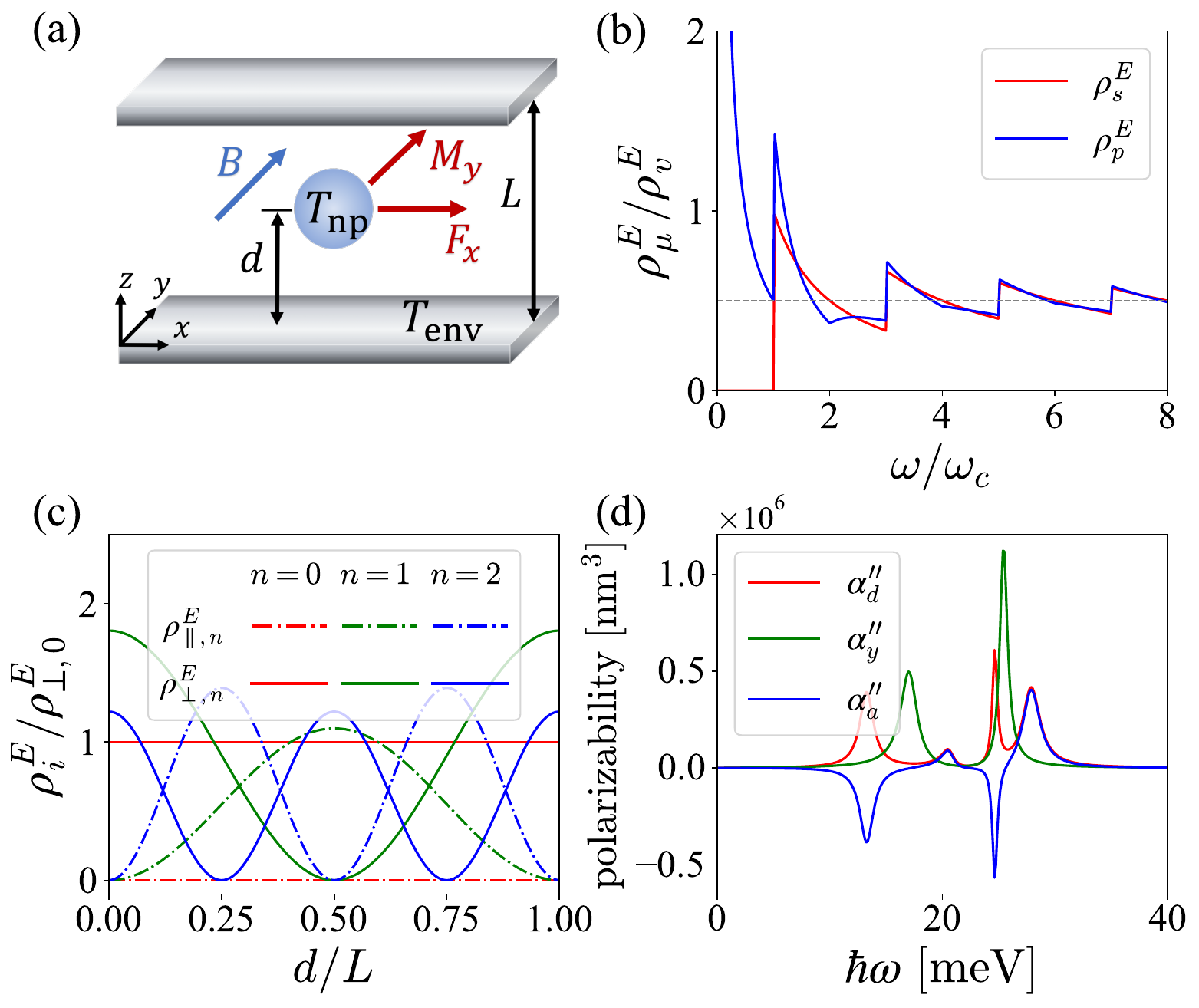} \\
\caption{(a) Schematic of a magneto-optic nanoparticle located at a distance $d$ above the
  lower mirror of a Fabry-P\'{e}rot cavity with length $L$. A temperature difference
  between the nanoparticle and the cavity environment drives radiative heat transfer, as
  well as a lateral force and a torque. With an external magnetic field $B$ applied along
  the $y$ direction, the resulting lateral force and torque are along the $x$ and $y$
  axes, respectively.
  (b) Normalized LDOS at the cavity center for $s$- and $p$-polarized modes as a function
  of the normalized frequency $\omega/\omega_c$ with $L = 60\,\mu$m.
  The normalizations are with respect to the vacuum electric density of states $\rho_v^E$.
  The cavity fundamental resonant frequency is $\hbar\omega_c =\hbar\pi c/L = 10.3\,$meV.
  (c) Normalized spatial distribution of the parallel and perpendicular electric LDOS at
  $\omega = 3.2 \omega_c$ for the lowest-order cavity modes. 
  The normalizations are with respect to $\rho^E_{\perp ,0}$.
  (d) The polarizability tensor components of the magneto-optic nanoparticle InSb with
  radius $R = 50\,$nm and magnetic field $B=4\,$T. 
}
\label{fig1}
\end{figure}

As schematically shown in Fig.~\ref{fig1}(a), the system under consideration consists of a
spherical nanoparticle of radius $R$ embedded within a Fabry-P\'{e}rot optical cavity. The
cavity is formed by two parallel perfect metallic plates separated by a distance $L$.
Before evaluating the radiative heat and momentum transfer between the nanoparticle and
its surrounding photonic environment, we first discuss the cavity properties.

In the frequency domain, the electric field $\bm{E}(\bm{r},\omega)$ at point $\bm{r}$
produced by a fluctuating current density $\bm{J}(\bm{r}', \omega)$ is given by the linear
response relation with
\begin{equation} \label{EGJ} 
  \bm{E}(\bm{r},\omega) = i\mu_0 \omega \int d\bm{r}' G(\bm{r},\bm{r}',\omega)
  \bm{J}(\bm{r}',\omega) .
\end{equation}
Due to the in-plane translational invariance, we express the dyadic Green's function via
Fourier transform:
\begin{equation} \label{G_FT}
  G(\bm{r}_1,\bm{r}_2)=\int \frac{d^2\bm{q}}{(2\pi)^2}  
  e^{i\bm{q}\cdot(\bm{R}_1-\bm{R}_2)} G(\bm{q}, z_1,z_2) ,
\end{equation}
where $\bm{r}=(\bm{R},z)$ separates the in-plane and out-of-plane coordinates. The
in-plane wavevector is denoted by $\bm{q}=(q_x, q_y)$, and the out-of-plane wavevector
component $\gamma_0$ satisfies the dispersion relation $\gamma_0^2+q^2=k_0^2$ with
$k_0=\omega/c$ the vacuum wavevector. 
Following Sipe's method in Ref.~\cite{sipe87}, the Green's function is decomposed into a
bulk vacuum contribution and a scattering term due to boundary reflections. The Green's
function within the cavity describing the field at $z_1$ generated by a source at $z_2$
with $z_1 \geq z_2$ is written as
\begin{align}
  G(\bm{q},z_1,z_2) =& \frac{i}{2\gamma_0}\sum_{\mu=s,p}
  \big[ e^{i\gamma_0(z_1-z_2)} (\tilde{r}_{\mu +} \hat{\mu}_+ 
  + \tilde{r}_{\mu -} \hat{\mu}_-) \hat{\mu}_+ \notag \\
  +& e^{i\gamma_0(z_1+z_2)} r_\mu(\tilde{r}_{\mu +} \hat{\mu}_+ 
  + \tilde{r}_{\mu -} \hat{\mu}_-)\hat{\mu}_-\big], \label{G_cavity}
\end{align}
where the multiple-reflection factors are given by
\begin{equation}
  \tilde{r}_{\mu +}=\frac{1}{1 - r_{\mu}^2 e^{2i \gamma_0 L}} , \qquad
  \tilde{r}_{\mu -}=\frac{r_\mu e^{2i\gamma_0(L-z_1)}}{1-r_\mu^2 e^{2i \gamma_0 L}} .
\end{equation}
Here, the subscript $\mu \in \{s,p\}$ denotes the polarization state, and $r_\mu$ is the
Fresnel reflection coefficient of the plates. Using dyadic notation, the unit
polarization vectors for upward ($+$) and downward ($-$) propagating waves are given by
\begin{equation} \label{sp}
  \hat{s}_{\pm} = \frac{1}{q} 
  \begin{bmatrix}
    q_y \\  -q_x \\ 0
  \end{bmatrix} , \qquad
  \hat{p}_{\pm} = 
  \frac{1}{qk_0} 
  \begin{bmatrix}
    \mp\gamma_0 q_x \\  \mp\gamma_0 q_y \\ q^2
  \end{bmatrix} .
\end{equation}
As a consistency check, this cavity Green's function reduces to the single-plate limit by
setting $e^{2i \gamma_0 L} \to 0$ under $L\to \infty$. 

In the limit of perfectly conducting mirrors with $r_s=-1$ and $r_p=1$, the Green's
function at $z_1=z_2=d$ reduces to
\begin{equation}
  G(\bm{q},d,d) = G_s(\bm{q},d,d) + G_p(\bm{q},d,d) ,
\end{equation}
with the contribution from the $s$-polarized mode
\begin{equation} \label{Gs}
  G_s(\bm{q},d,d) = \frac{i\big[1 - e^{2i\gamma_0(L-d)}\big](1 - e^{2i\gamma_0 d})
  \hat{s}\hat{s}}{2\gamma_0 (1 - e^{2i\gamma_0 L})}  ,
\end{equation}
and the one from the $p$-polarized mode
\begin{equation} \label{Gp}
  G_p(\bm{q},d,d) = \frac{i\big[ \hat{p}_+ + e^{2i\gamma_0(L-d)} \hat{p}_- \big] \big(
  \hat{p}_+ + e^{2i\gamma_0 d} \hat{p}_- \big)}{2\gamma_0 (1 - e^{2i\gamma_0 L})} . 
\end{equation}

The electric LDOS inside the cavity is given by~\cite{Joulain03}
\begin{equation}\label{rhoE}
  \rho^E(d, \omega) = \frac{\omega}{\pi c^2} 
  \int \frac{d^2\bm{q}}{(2\pi)^2} {\rm Im} \{ {\rm Tr} [ G(\bm{q}, d, d) ] \},
\end{equation}
which can be decomposed into $s$- and $p$-polarized contributions with $\rho^E
= \rho_s^E + \rho_p^E$. The $s$-polarized electric LDOS is expressed as
\begin{equation}
  \rho_s^E(d,\omega) = \int_0^{k_0} \frac{qdq}{(2\pi)^2} \frac{k_0}{c\gamma_0}{\rm Re}
  \bigg[ \frac{1 - e^{2i\gamma_0(L-d)}}{1-e^{2i\gamma_0L}} (1 - e^{2i\gamma_0 d}) \bigg],
\end{equation}
and the $p$-polarized one is
\begin{align}
  \rho_p^E(d,\omega) = \int_0^{k_0} & \frac{qdq}{(2\pi)^2} \frac{k_0}{c\gamma_0} 
  {\rm Re}\Big\{ \frac{1}{1-e^{2i\gamma_0L}} \Big[1+e^{2i\gamma_0L} +  \notag\\
  & \Big(\frac{2q^2}{k_0^2}-1\Big) \Big( e^{2i\gamma_0d}
  + e^{2i\gamma_0(L-d)} \Big) \Big] \Big\} ,
\end{align}
where we have expressed the integral in polar coordinate system and used the relations
${\rm Tr}(\hat{s}\hat{s}) = {\rm Tr}(\hat{p}_+ \hat{p}_+) = {\rm Tr}(\hat{p}_- \hat{p}_-)
= 1$ and ${\rm Tr}(\hat{p}_+ \hat{p}_-) = {\rm Tr}(\hat{p}_- \hat{p}_+) = 2q^2/k_0^2 - 1$.
Notably, the near-field evanescent modes ($q>k_0$) make no contribution since in this
regime the purely imaginary out-of-plane wavevector $\gamma_0$ causes the dyadic Green's
functions in Eqs.~\eqref{Gs} and \eqref{Gp} to be real.
We use contour integration to calculate the above integrals.
By applying the residue theorem at the poles $\gamma_0 = n\pi/L$ (for $n \ge 1$) and
evaluating the contribution from the quarter-circular arc around the origin ($n=0$), we
obtain the $s$-polarized electric LDOS as
\begin{equation}
  \rho_s^E(d,\omega)
  = \frac{k_0}{2\pi cL}\sum_{n=1}^{n_c}\sin^2\Big(\frac{n\pi d}{L}\Big) ,
\end{equation}
where we have defined the fundamental resonant frequency $\omega_c=\pi c/L$ and the mode
cutoff index $n_c\equiv\lfloor \omega/\omega_c \rfloor$. 
Similarly, the $p$-polarized electric LDOS is obtained as
\begin{align}
  & \rho_p^E(d,\omega) = \rho^E_{\perp ,0}(d,\omega) + \notag \\
  &\ \ \frac{k_0}{2\pi cL}\sum_{n=1}^{n_c}\bigg[ \cos^2\Big(\frac{n\pi d}{L}\Big)
  - \Big(\frac{n\omega_c}{\omega}\Big)^2 \cos\Big(\frac{2n\pi d}{L}\Big)\bigg], 
\end{align}
where the zeroth-order mode is given by
\begin{equation} \label{perp_n0}
  \rho^E_{\perp ,0}(d,\omega) = \frac{k_0}{4\pi cL} .
\end{equation}
Summing these polarization-resolved contributions yields the total electric LDOS with
\begin{equation}
  \rho^E(d,\omega) =\frac{k_0}{4\pi cL}+\frac{k_0}{2\pi cL}\sum_{n=1}^{n_c}\bigg[ 1 -
  \Big(\frac{n\omega_c}{\omega}\Big)^2 \cos \Big(\frac{2n\pi d}{L}\Big)\bigg] .
\end{equation}

To visualize these cavity properties, Fig.~\ref{fig1}(b) displays the normalized electric
LDOS for $s$- and $p$-polarized modes at the cavity center. The normalizations
are with respect to the vacuum electric density of states, which is half of the total
vacuum density of states given by $\rho_v^E(\omega)= \omega^2 / (2\pi^2c^3)$.
The $s$-polarized LDOS vanishes below the fundamental resonant frequency of the cavity
$\omega_c$. This complete suppression occurs because it possesses purely in-plane electric
fields, which are forced to zero at the perfectly conducting boundaries. Consequently, a
spatially constant zeroth-order mode is forbidden, and there is no $s$-polarized states
below the $n=1$ cutoff threshold.
In contrast, the normalized $p$-polarized LDOS diverge in the low-frequency limit. This
divergence originates from the zeroth-order contribution which scales linearly with
frequency. Thus, normalizing it against the vacuum density of states with scaling
$\omega^2$ yields a $1/\omega$ divergence as $\omega \to 0$. It is noteworthy, however,
that the unnormalized $p$-polarized LDOS vanishes as the frequency approaches zero. 
The characteristic stepwise enhancements at $\omega = n\omega_c$ with $n\geq 1$ correspond
to the activation of successive cavity modes as the frequency surpasses each discrete
cutoff threshold.

Alternatively, we can decompose the electric LDOS into components parallel and
perpendicular to the cavity mirrors as in Ref.~\cite{cavity-25}. Due to the in-plane
rotational symmetry, the parallel electric LDOS can be obtained from either transverse
component ($G_{xx}$ or $G_{yy}$) of the dyadic Green's function:
\begin{align}
  \rho^E_{\parallel}(d,\omega) 
  =& \frac{\omega}{\pi c^2} \int \frac{d^2\bm{q}}{(2\pi)^2} {\rm Im}[G_{xx}(\bm{q},d,d)]
  \notag \\
  =& \sum_{n=0}^{n_c} \rho^E_{\parallel ,n}(d,\omega) ,
\end{align}
where the contribution from the $n$-th order cavity mode is explicitly given by
\begin{equation}
  \rho^E_{\parallel ,n}(d,\omega) = \frac{k_0}{4\pi cL} \bigg[ 1 + \Big(
  \frac{n\omega_c}{\omega}\Big)^2 \bigg] \sin^2\Big(\frac{n\pi d}{L}\Big) .
\end{equation}
Notably, the zeroth-order mode vanishes, i.e., $\rho^E_{\parallel ,0}(d,\omega) =0$ due to
the perfectly conducting boundaries. Similarly, the perpendicular electric LDOS is
obtained from the longitudinal component with
\begin{align}
  \rho^E_{\perp}(d,\omega) 
  =& \frac{\omega}{\pi c^2} \int \frac{d^2\bm{q}}{(2\pi)^2} {\rm Im}[G_{zz}(\bm{q}, d, d)] 
  \notag \\
  =& \sum_{n=0}^{n_c} \rho^E_{\perp ,n}(d,\omega) ,
\end{align}
where the zeroth-order mode is given by Eq.~\eqref{perp_n0},
and the contributions from the higher-order modes ($n\geq 1$) take the form
\begin{equation}
  \rho^E_{\perp ,n}(d,\omega) 
  = \frac{k_0}{2\pi cL} \bigg[1-\Big(\frac{n\omega_c}{\omega}
  \Big)^2\bigg]\cos^2\Big(\frac{n\pi d}{L} \Big) .
\end{equation}
The total electric LDOS is recovered by summing the spatially resolved contributions with
\begin{equation}
  \rho^E(d,\omega) = 2\rho_{\parallel}^E(d,\omega) + \rho_{\perp}^E(d,\omega) .
\end{equation}

Figure~\ref{fig1}(c) shows the normalized spatial distribution of the parallel
($\rho^E_{\parallel ,n}$) and perpendicular ($\rho^E_{\perp ,n}$) electric LDOS for the
lowest-order cavity modes with $n=0,1,2$. 
The normalizations are with respect to $\rho^E_{\perp ,0}$ given in Eq.~\eqref{perp_n0}.
Due to the perfectly conducting boundary conditions, the parallel LDOS vanishes at the
mirror surfaces for all modes, and is absent for the zeroth-order mode. In contrast, the
perpendicular LDOS is maximized at these boundaries. At the cavity center, the
higher-order modes ($n\geq 1$) exhibit distinct spatial parity: the odd mode possesses a
parallel LDOS antinode and a perpendicular node, whereas the even mode behaves oppositely.

\section{Radiative heat and momentum transfer} \label{sec:III}
In this section, we first derive general expressions for radiative heat, linear momentum,
and angular momentum transfer between the magneto-optic nanoparticle and the cavity
environment. According to the fluctuation-dissipation theorem, the symmetrized correlation
function of the fluctuating electric field $\bm{E}^{\rm fl}$ from the photonic environment
is given by~\cite{review05}
\begin{align}
  \big< E_j^{\rm fl}(\bm{r}, & \omega) E_k^{\rm fl*}(\bm{r}',\omega')\big>=
  \frac{\mu_0\hbar \omega^2}{2i} \bigg[ N_{\rm env}(\omega) + \frac{1}{2} \bigg] \times
  \notag \\
  & \big[ G_{jk}(\bm{r},\bm{r}',\omega)-G_{kj}^*(\bm{r}',\bm{r},\omega) \big]
  4\pi\delta(\omega-\omega') , \label{EjEk}
\end{align}
where the subscripts $j$ and $k$ denote Cartesian indices, and $N_{\rm env}(\omega) =
[\exp(\hbar\omega/k_B T_{\rm env})-1]^{-1}$ represents the Bose-Einstein distribution at
the environmental temperature $T_{\rm env}$.
The nanoparticle, maintained at a temperature $T_{\rm np}$, hosts thermally fluctuating
dipole moments $\bm{p}^{\rm fl}$ whose symmetrized correlations satisfy
\begin{equation} \label{pjpk}
  \langle p_j^{\rm fl}(\omega)p_k^{\rm fl*}(\omega')\rangle
  =\frac{\hbar\epsilon_0}{2i}(\alpha_{jk}-\alpha_{kj}^*) 
  \bigg[ N_{\rm np}(\omega) + \frac{1}{2} \bigg] 4\pi\delta(\omega-\omega'),
\end{equation}
with $N_{\rm np}(\omega) = [\exp(\hbar\omega/k_B T_{\rm np})-1]^{-1}$.
The interaction between the nanoparticle and the surrounding photonic environment involves
mutual electromagnetic induction. The fluctuating environmental electric field induces a
local dipole moment in the nanoparticle, 
\begin{equation} \label{p_in}
 p_j^{\rm in} = \epsilon_0 \sum_k\alpha_{jk} E_k^{\rm fl} ,
\end{equation}
with $\alpha$ the nanoparticle's polarizability tensor. Meanwhile, the fluctuating dipole
moment of the nanoparticle at position $\bm{r}_1$ induces an electric field $\bm{E}^{\rm
in}$ at position $\bm{r}_2$ with
\begin{equation} \label{E_in}
  E_j^{\rm in}(\bm{r}_2,\omega)=\mu_0\omega^2\sum_k G_{jk}(\bm{r}_2,\bm{r}_1,\omega)
  p_k^{\rm fl}(\bm{r}_1,\omega).
\end{equation}

Within the framework of fluctuational electrodynamics, the total exchanged power $H$,
lateral force $F_j$, and torque $M_j$ with $j=x,y,z$ acting on the nanoparticle arise from
the cross-correlations between the fluctuating sources and their respective induced
responses. Physical quantities $\mathsf{X} \in \{H, F_j, M_j\}$ are evaluated by
integrating their corresponding spectral densities $\mathsf{x} \in \{h, f_j, m_j\}$ over
frequencies:
\begin{equation}
 \mathsf{X} = \int_0^{\infty} \frac{d\omega}{2\pi} \mathsf{x}(\omega) .
\end{equation}
The spectral densities for the radiative power, lateral force, and torque can be
respectively expressed as~\cite{lateral_21-2}
\begin{align}
  h(\omega) &= 2\sum_k \int_{\omega'} \omega {\rm Im}\big[\big\langle 
  p_k^{\rm fl*}(\omega)E_k^{\rm in}(\omega') \big\rangle 
  +({\rm in} \leftrightarrow {\rm fl}) \big],  \label{Eqh} \\
  f_j(\omega)&= 2\sum_k\int_{\omega'} {\rm Re}\big[\big\langle 
  p_k^{\rm fl}(\omega) \partial_j E_k^{\rm in*}(\omega')\big\rangle
  +({\rm in} \leftrightarrow {\rm fl}) \big],  \label{Eqf} \\
  m_j(\omega)&= 2\sum_{kl}\int_{\omega'} \epsilon_{jkl} {\rm Re}\big[\big\langle 
  p_k^{\rm fl}(\omega) E_l^{{\rm in}*}(\omega')\big\rangle
  +({\rm in} \leftrightarrow {\rm fl}) \big],  \label{Eqm}
\end{align}
where $\epsilon_{jkl}$ is the Levi-Civita symbol, and the abbreviated frequency
integration is defined as 
\begin{equation*}
  \int_{\omega'} \equiv \int_{-\infty}^{\infty} \frac{d\omega'}{2\pi}
  e^{i(\omega-\omega')t} .
\end{equation*}

By combining the fluctuation-dissipation theorems with Eqs.~\eqref{Eqh}-\eqref{Eqm} and
substituting the Green's function in Eq.~\eqref{G_cavity}, we can obtain expressions for
the spectral densities. Details are given in Appendix~\ref{AppendixB}. The power spectrum
takes the form
\begin{align}
  h(\omega) =& \hbar\omega k_0^2 N_d(\omega)\int_0^{\infty} \frac{qdq}{2\pi}
  \Big\{ (\alpha_d''+\alpha_y''){\rm Re}\Big[ (1 + r_s e^{2i\gamma_0 d}) \notag \\
  & (\tilde{r}_{s+}+\tilde{r}_{s-})\frac{1}{\gamma_0} +
  (1-r_pe^{2i\gamma_0d})(\tilde{r}_{p+}-\tilde{r}_{p-}) \frac{\gamma_0}{k_0^2}\Big]
  \notag\\ 
  +& \alpha_d''{\rm Re}\Big[ (1+r_pe^{2i\gamma_0d})(\tilde{r}_{p+}+\tilde{r}_{p-})
  \frac{2q^2}{\gamma_0k_0^2} \Big] \Big\}, \label{h_cavity}
\end{align}
where $\alpha_d$, $\alpha_y$, and $\alpha_a$ represent the transverse diagonal,
longitudinal diagonal, and off-diagonal components of the polarizability
tensor, respectively, as shown in Appendix~\ref{AppendixA}. 
The Bose-Einstein distribution difference is given by
\begin{equation}
  N_d(\omega)=N_{\rm np}(\omega)-N_{\rm env}(\omega) .
\end{equation}
By applying the magnetic field along the $y$ direction, the induced magneto-optical 
anisotropy breaks the spatial reflection symmetry in the $x-z$ plane. The resulting 
off-diagonal polarizability components $\pm i\alpha_a''$ couple the $x$- and $z$-directed 
field fluctuations. This nonreciprocal cross-coupling skews the scattered wavevector 
spectrum, which directly drives the lateral force $f_x$. 
Simultaneously, the imaginary unit $i$ in these off-diagonal components dictates a $\pi/4$
phase shift between the orthogonal dipole responses. Because of this phase delay, the
thermally excited dipoles rotate within the $x-z$ plane so that the nanoparticle emits and
absorbs elliptically polarized thermal radiation, which carries spin angular momentum. The
temporal average of the electromagnetic angular momentum transfer $\bm{p} \times \bm{E}$
is thus directed along the $y$ axis, leading to the macroscopic torque $m_y$.
While the out-of-plane spatial gradient of the cavity field inherently induces a
conventional Casimir force $f_z$, we omit its discussion here to focus on the lateral
force due to the nonreciprocity.
The spectral density of the lateral force is obtained as
\begin{equation} \label{f_cavity}
  f_x(\omega)=\hbar N_d(\omega)\alpha_a''\int_0^{\infty}\frac{q^3dq}{\pi}
  {\rm Im}\big( r_p \tilde{r}_{p+}e^{2i\gamma_0 d}-\tilde{r}_{p-} \big) ,
\end{equation}
and the corresponding spectral density of the torque yields
\begin{align}
  m_y(\omega) =& -\hbar k_0^2N_d(\omega)\alpha_a''\int_0^{\infty}\frac{qdq}{2\pi} 
  {\rm Re}\bigg[ \frac{1}{\gamma_0}(1+r_se^{2i\gamma_0d})  \notag\\
  &(\tilde{r}_{s+}+\tilde{r}_{s-}) +
  \frac{\gamma_0}{k_0^2}(1-r_pe^{2i\gamma_0d})(\tilde{r}_{p+}-\tilde{r}_{p-}) \notag\\ 
  &+(1+r_pe^{2i\gamma_0d})(\tilde{r}_{p+}+\tilde{r}_{p-}) \frac{2q^2}{\gamma_0k_0^2}\bigg]
  . \label{m_cavity}
\end{align}
These expressions indicate the following two features.
First, the heat transfer is governed  by the diagonal components of the polarizability
tensor, whereas the momentum transfer is proportional to the off-diagonal components
responsible for the system's nonreciprocity. 
Second, while both $s$- and $p$-polarized modes mediate the transfer of heat and angular
momentum, the lateral force arises only from the $p$-polarized mode. Physically, as shown
in Eq.~\eqref{sp}, the electric field of an $s$-polarized mode is parallel to the mirror
surfaces, lacking an out-of-plane component. Consequently, it is unable to couple to the
symmetry-broken dipole transitions between the $x$ and $z$ axes, which is required to
generate a lateral force along the $x$ direction. 

In the following, we derive explicit expressions for the spectral densities in the limit
of perfectly conducting mirrors and then in three geometric regimes: the single-plate
limit ($L \to \infty$), the near-surface regime ($d/L \to 0$), and the cavity center ($d =
L/2$), followed by interpretation of the numerical results. 
The numerical calculations assume a nanoparticle temperature of $T_{\rm np}=330\,$K and an
environmental temperature of $T_{\rm env}=300\,$K. The radius of the nanoparticle is $R =
50\,$nm and the magnetic field applied to it is $B=4\,$T, which corresponds to the
polarizability tensor components illustrated in Fig.~\ref{fig1}(d). Spectrally aligning
with the thermal energy scale ($k_B T_{\rm np}$), the polarizability resonances
effectively mediate the energy and momentum transfer in this work.

\subsection{Radiative heat transfer}
In the limit of perfectly conducting mirrors with $r_s = -1$ and $r_p = 1$, the power
spectrum in Eq.~\eqref{h_cavity} simplifies to summation over the discrete cavity modes:
\begin{align}  \label{h0}
  h(\omega)= 4\pi \hbar\omega^2 N_d(\omega) \sum_{n=0}^{n_c} \big[ 
  & (\alpha_d''+\alpha_y'') \rho^E_{\parallel,n}(d,\omega) \notag \\ 
  +& \alpha_d'' \rho^E_{\perp,n}(d,\omega) \big] ,
\end{align}
which is consistent with the results obtained in Ref.~\cite{cavity-25}. 
Below, we discuss several limits of this expression of the power spectrum. First, in the
single-plate limit with $L\to \infty$, the discrete cavity modes merge into a continuous
spectrum. By introducing the continuous out-of-plane wavevector $\gamma = n\pi/L$, the
mode spacing becomes infinitesimal, $\Delta\gamma = \pi/L \to d\gamma$. Replacing the
discrete sum with an integral over $\gamma$, Eq.~\eqref{h0} transforms into
\begin{align}
  h(\omega)=\hbar\omega k_0^2N_d(\omega) & \int_0^{k_0} \frac{d\gamma}{\pi} 
  \bigg[(\alpha_d'' +\alpha_y'')\Big(1+\frac{\gamma^2}{k_0^2}\Big) \sin^2(\gamma d) 
  \notag\\
  &+ 2\alpha_d'' \bigg(1-\frac{\gamma^2}{k_0^2}\bigg) \cos^2(\gamma d) \bigg] .
\end{align}
Evaluating this integral reduces the expression to the single-plate formula
\begin{align}
  h(\omega)= \hbar\omega k_0^3 N_d(\omega)&\big\{(\alpha_d''+\alpha_y'')
  \big[1+\mathcal{S}(\beta)-3\sin\beta / \beta \big] \notag\\
  &+2\alpha_d'' \mathcal{S}(\beta) \big\} / (3\pi) , \label{h1}
\end{align}
with the dimensionless parameter $\beta=2k_0 d$ and the function $\mathcal{S}(\beta)= 1 +
3(\sin\beta-\beta\cos\beta)/\beta^3$. The parameter $\beta$ represents the round-trip
phase accumulated by the electromagnetic field between the nanoparticle and the mirror.
Consequently, the oscillatory terms in Eq.~(\ref{h1}) originate from the interference
between the direct and the reflected fields. In the limit $d\to \infty$, the interference
vanishes, $\mathcal{S}(\beta)\to 1$ and $\sin\beta/\beta\to 0$, reducing Eq.~(\ref{h1}) to
the free-space result
$h(\omega) = 2\hbar\omega k_0^3N_d(\omega)(2\alpha_d''+\alpha_y'') / (3\pi)$.
In the near-surface limit $\beta\to 0$, one finds
$\mathcal S(\beta)\to 2$ and $\sin\beta/\beta \to 1$, yielding $h(\omega)\to 4\hbar\omega
k_0^3N_d(\omega)\alpha_d''/(3\pi)$. Notably, the contribution associated with
$\alpha_y''$ disappears, reflecting the suppression of tangential electric-field
fluctuations at a perfectly conducting boundary, such that only the dipole component normal
to the surface participates in the radiative heat exchange.

Second, returning to the finite-cavity configuration, we consider the near-surface regime
where the nanoparticle approaches the lower mirror with $d \to 0$. In this limit, the
power spectrum takes the form 
\begin{equation} \label{h2}
  h(\omega) = \hbar\omega k_0^2 N_d(\omega) \frac{\alpha_d''}{L}
  \bigg\{1+\sum_{n=1}^{n_c}
  2\bigg[1-\Big(\frac{n\omega_c}{\omega}\Big)^2\bigg]\bigg\}. 
\end{equation}
Finally, when the nanoparticle is positioned at the cavity center with $d = L/2$, the
spatial symmetry of the Fabry-P\'{e}rot modes decouples the radiative heat transfer into
distinct even and odd modal contributions, reducing the power spectrum to
\begin{align}
  h(\omega) = \hbar\omega & k_0^2 N_d(\omega) \Bigg\{ \sum_{n \in \{1, 3 \dots\}}^{n_c}
  \frac{\alpha_d''+\alpha_y''}{L} \bigg[1+ \Big(\frac{n\omega_c}{\omega}\Big)^2\bigg]
  \notag \\
  &+ \frac{\alpha_d''}{L} + \sum_{n \in \{2, 4 \dots\}}^{n_c} \frac{2\alpha_d''}{L}
  \bigg[1-\Big(\frac{n\omega_c}{\omega}\Big)^2\bigg] \Bigg\} . \label{h3}
\end{align}
Physically, this indicates that the odd cavity modes couple to the parallel components of
the nanoparticle's polarizability ($\alpha_d''$ and $\alpha_y''$) due to the presence of
electric-field antinodes in the transverse plane, whereas the even modes couple to the
perpendicular polarizability component ($\alpha_d''$).
In the sub-wavelength regime with $L \ll \lambda_{\rm th}$ where the characteristic
thermal wavelength is defined as $\lambda_{\rm th} = \hbar c / k_B T$, both
Eqs.~\eqref{h2} and \eqref{h3} reduce to the same limit, given by
\begin{equation} \label{h4}
  h(\omega) = \hbar\omega k_0^2 N_d(\omega) \alpha_d'' / L .
\end{equation}
At room temperature, the thermal wavelength $\lambda_{\rm th}$ is approximately
$7.6\,\mu$m. This expression indicates that the exchanged power diverges as $1/L$
due to the spatial squeezing of the fundamental vacuum field. However, it is noteworthy
that this divergence is physically regularized at extremely short distances by the finite
size of the nanoparticle, since this work relies on the electric-dipole approximation in
which the nanoparticle radius $R$ is much smaller than both the particle-to-mirror
distance and the thermal wavelength ($R \ll d, \lambda_{\rm th}$).

\begin{figure}
\centering
\includegraphics[width=\columnwidth]{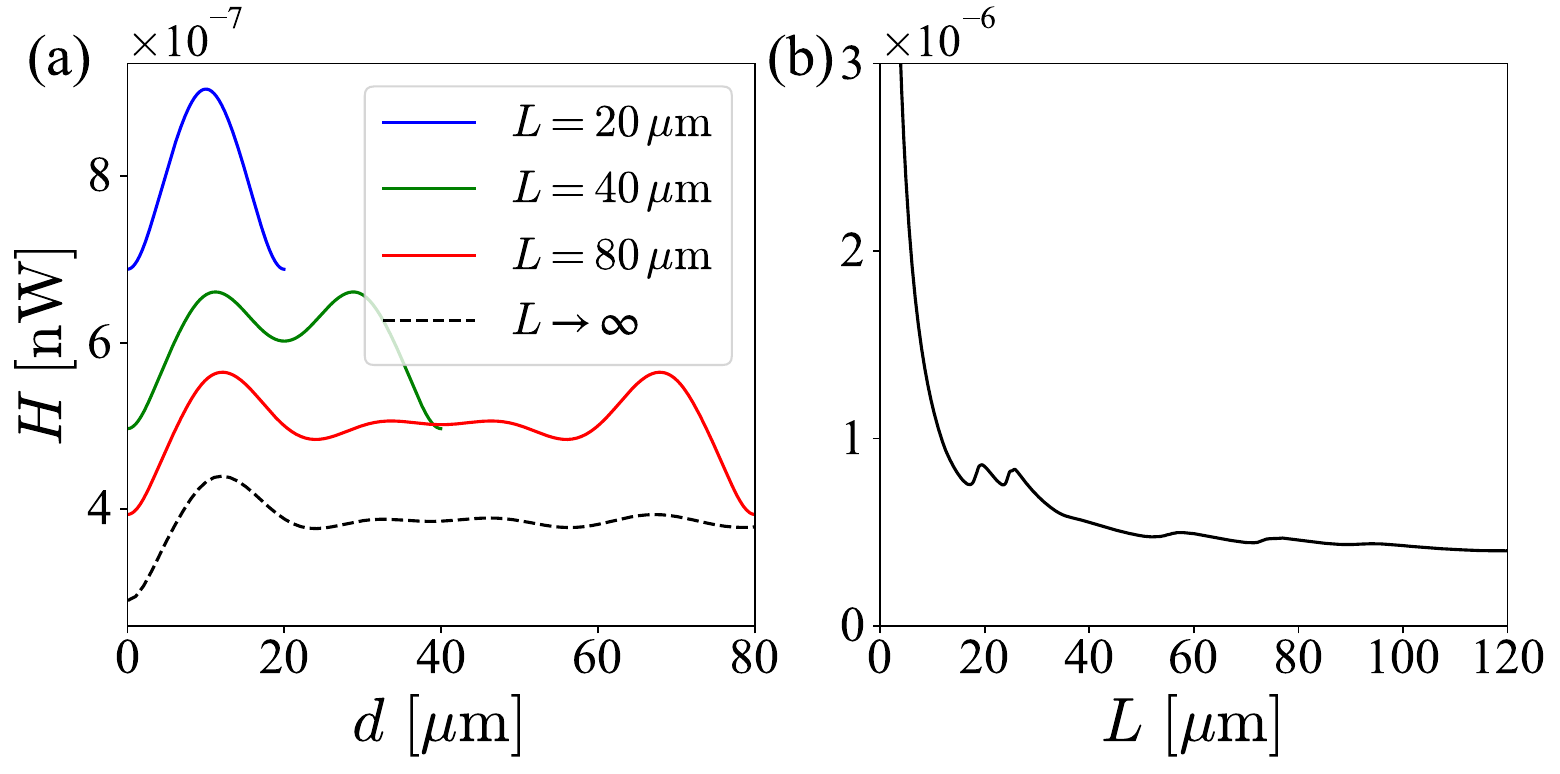} \\
\caption{(a) Exchanged power as a function of the particle-to-mirror distance $d$ for
various cavity lengths $L$. (b) Exchanged power at the cavity center as a function of $L$.
}
\label{fig2}
\end{figure}

Figure~\ref{fig2}(a) shows the exchanged power $H$ as a function of the particle position
$d$. Compared with the single-plate limit (dashed line), the Fabry-P\'{e}rot cavity
substantially enhances the radiative heat transfer, demonstrating the thermal Purcell
effect. This enhancement originates from the cavity-induced modification of the LDOS,
which increases the number of thermally accessible photonic states available for energy
exchange. The effect becomes stronger as the cavity length decreases since the
electromagnetic field is confined within a smaller volume. In addition to the overall
enhancement, the exchanged power exhibits pronounced oscillations as the
particle-to-mirror distance $d$ varies. These oscillations reflect the standing-wave
structure of the cavity modes. As the cavity length increases, more cavity modes
contribute within the thermal spectral window, producing more complex interference
patterns.

The influence of geometric confinement is further illustrated in Fig.~\ref{fig2}(b), which
shows the exchanged power at the cavity center as a function of the cavity length $L$. For
$L \ll \lambda_{\rm th}$, the heat transfer increases approximately as $1/L$, consistent
with the asymptotic limit in Eq.~\eqref{h4}. In this regime, all higher-order modes are
cut off, and the cavity supports only the zeroth-order mode. Because this mode features a
uniform out-of-plane electric field, the strong spatial confinement increases the
perpendicular LDOS given by Eq.~\eqref{perp_n0}, resulting in an enhanced power exchange.

As the cavity length becomes comparable with the thermal wavelength, the exchanged power
develops oscillatory features associated with the successive activation of higher-order
cavity modes. At the cavity center, the parity selection rule in Eq.~\eqref{h3} separates
odd and even modes into different coupling channels. Odd modes couple to the parallel
electric-field fluctuations with a frequency coefficient of $(1+n^2\omega_c^2/\omega^2)$.
When an odd mode turns on ($\omega = n\omega_c$), this coefficient evaluates to $2$,
resulting in a finite discontinuity in the LDOS. 
Conversely, even modes couple to the perpendicular fluctuations with a coefficient of
$(1-n^2\omega_c^2/\omega^2)$. Because this term evaluates to zero at the cutoff frequency,
even modes smoothly enter the spectrum without inducing abrupt changes. The integration of
these step-wise modal activations over the broadband thermal spectrum produces the
oscillatory behavior in Fig.~\ref{fig2}(b).

Finally, for sufficiently large cavity lengths with $L\gg\lambda_{\rm th}$, the mode
spacing becomes negligible, the discrete spectrum approaches a continuum, and the
exchanged power gradually converges to the single-plate limit.

\subsection{Radiative linear-momentum transfer}
In the limit of perfectly conducting mirrors, the lateral force spectrum in
Eq.~\eqref{f_cavity} simplifies to
\begin{equation} \label{f0}
  f_x(\omega)=\hbar k_0^2N_d(\omega)\alpha_a''\sum_{n=0}^{n_c}\frac{n\pi}{L^2}\bigg[
  1-\Big(\frac{n\omega_c}{\omega}\Big)^2\bigg]\sin\Big(\frac{2n\pi d}{L}\Big).
\end{equation}
To obtain the single-plate limit, we use the same procedure used to derive Eq.~\eqref{h1}.
By merging the discrete cavity modes into a continuous spectrum and integrating over the
out-of-plane wavevector $\gamma$, the lateral force spectrum reduces to
\begin{equation} \label{f1}
  f_x(\omega) = \frac{\hbar}{8\pi d^4} N_d(\omega) \alpha_a''
  \big[ (3-\beta^2)\sin\beta - 3\beta\cos\beta \big] ,
\end{equation}
with $\beta=2k_0 d$.
This expression recovers the result obtained in Ref.~\cite{lateral_25} by setting the
environment temperature $T_{\rm env}$ to be zero.
Returning to the finite cavity geometry, Eq.~\eqref{f0} shows that the lateral force
vanishes at both the boundary ($d \to 0$) and the cavity center ($d = L/2$).
Physically, the lateral force is absent at the mirror surface because the perfectly
conducting boundary conditions force the parallel electric-field components to vanish.
At the cavity center, the spatial symmetry of the Fabry-P\'{e}rot mode structure prevents
asymmetric lateral momentum transfer.
The vanishing of the lateral force at $d \to 0$ and $d = L/2$ can also be understood
through the parity decomposition of the cavity modes. At these points, specific orthogonal
components of the photonic field vanish, thereby completely suppressing the cross-coupling
required for lateral momentum transfer.

\begin{figure}
\centering
\includegraphics[width=\columnwidth]{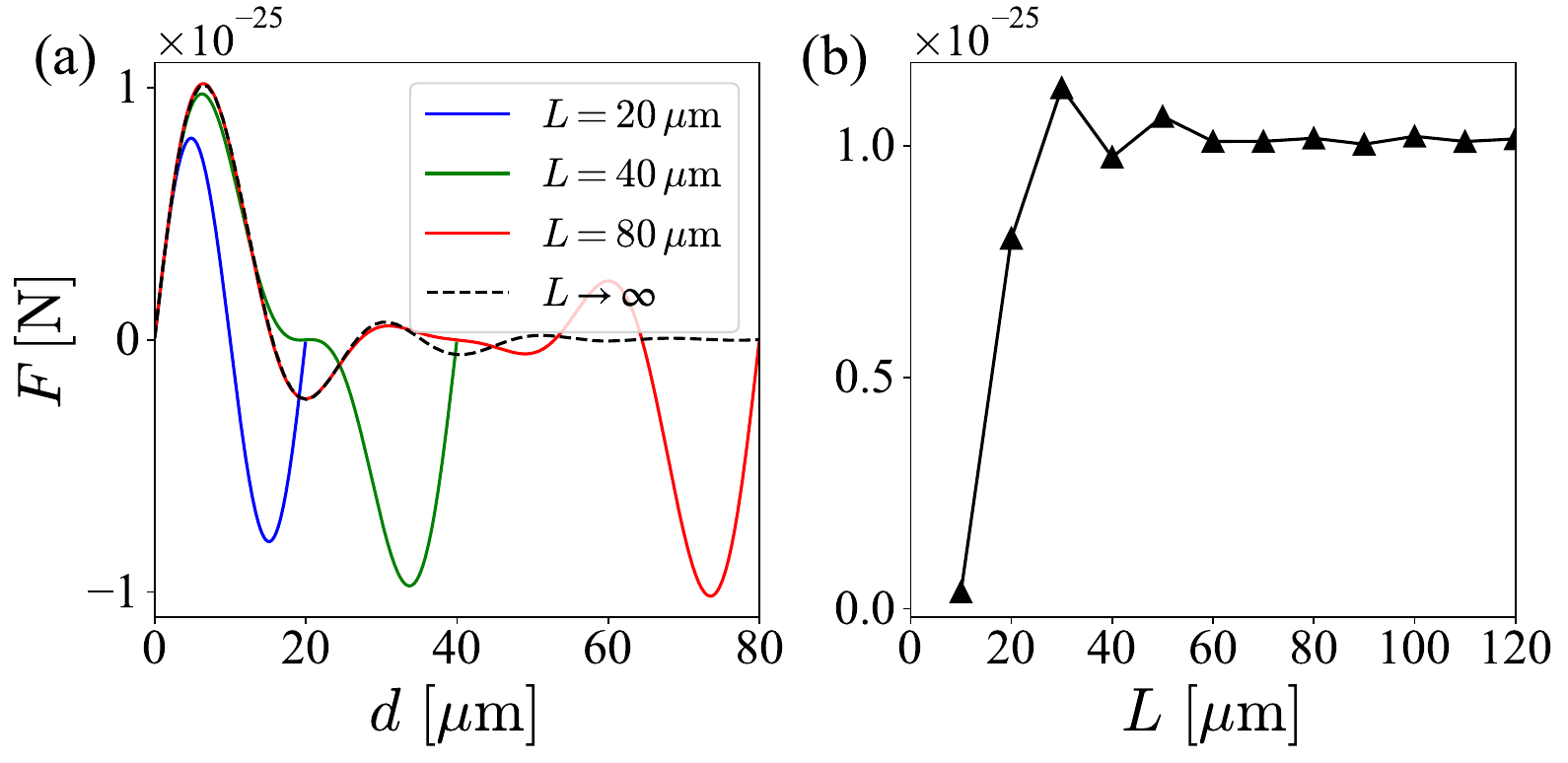} \\
\caption{(a) Lateral force acting on the nanoparticle as a function of the
  particle-to-mirror distance $d$ for various cavity lengths $L$. 
  (b) Maximum lateral force obtained by sweeping $d$, shown as a function of $L$ with a
  discrete step size of $10\mu$m. }
\label{fig3}
\end{figure}

The lateral force exhibits pronounced oscillatory behavior as a function of the particle
position, as shown in Fig.~\ref{fig3}(a). This oscillation directly follows from the
$\sin(2n\pi d/L)$ dependence in Eq.~\eqref{f0}. Unlike the exchanged power, which is
determined by the local electromagnetic energy density, the lateral force originates from
an asymmetric redistribution of momentum carried by thermally excited photons, and as a
consequence, it changes sign whenever the direction of the net momentum flux reverses.

For short cavity lengths, the geometric boundaries strongly modulate the magnitude of the
lateral force. This is demonstrated in Fig.~\ref{fig3}(b), which shows the maximum force,
obtained by sweeping the particle position $d$ for each cavity length $L$.
In contrast to the cavity-enhanced heat transfer shown in Fig.~\ref{fig2}, the results
demonstrate that geometric confinement generally suppresses linear-momentum transfer.
When the cavity length is small compared with the thermal wavelength, only a limited
number of cavity modes remain thermally accessible within the relevant spectral range.
Since the lateral force requires coupling between transverse and longitudinal fields, this
reduction in available modes drastically weakens the net momentum asymmetry. As the cavity
length increases, higher-order cavity modes become available, and the force gradually
increases. For relatively large cavities, the mode spacing becomes sufficiently small that
the discrete spectrum effectively recovers the continuum limit, and the force converges
toward the single-plate result.

\subsection{Radiative angular-momentum transfer}
In the limit of perfectly conducting mirrors, the torque spectrum in Eq.~\eqref{m_cavity}
simplifies to
\begin{align}
  &m_y(\omega)=-\hbar k_0^2N_d(\omega)\frac{\alpha_a''}{L}-\hbar k_0^2N_d(\omega)
  \frac{\alpha_a''}{L}\sum_{n=1}^{n_c}\notag\\
  &\bigg \{1+\Big(\frac{n \omega_c}{\omega}\Big)^2  
  +\bigg[1-3\Big(\frac{n \omega_c}{\omega}\Big)^2\bigg]
  \cos^2\Big(\frac{n\pi d}{L}\Big)\bigg\}.  \label{m0}
\end{align}
Following the same procedure applied previously, evaluating the single-plate limit with
$L\to \infty$ reduces it to
\begin{equation} \label{m1}
  m_y(\omega)=\frac{-\hbar \alpha_a''}{8\pi d^3}N_d(\omega)\big[4\beta^3 / 3
  +(3-\beta^2)\sin\beta-3\beta\cos\beta\big] .
\end{equation}
This expression approaches the free-space result $m_y(\omega) \to -4\hbar
k_0^3N_d(\omega)\alpha_a''/(3\pi)$ in both the limit $d\to \infty$ and the
near-surface limit $\beta\to 0$~\cite{nonrecip_rotate19, lateral_23-2}.

We now consider the finite-cavity configuration. When the nanoparticle is located near the
surface with $d \to 0$, the spectrum becomes
\begin{equation} \label{m2}
  m_y(\omega)=-\hbar k_0^2 N_d(\omega) \frac{\alpha_a''}{L} \bigg\{1+ \sum_{n=1}^{n_c}
  2\bigg[1-\Big(\frac{n\omega_c}{\omega}\Big)^2 \bigg] \bigg\} .
\end{equation}
When the nanoparticle is positioned at the cavity center with $d=L/2$, the torque spectrum
can be decoupled into even and odd modal contributions with
\begin{align}
  m_y(\omega) = -\hbar k_0^2 N_d(\omega) & \frac{\alpha_a''}{L} \Bigg\{1 + \sum_{n \in
  \{2, 4 \dots\}}^{n_c} 2\bigg[1-\Big(\frac{n\omega_c}{\omega}\Big)^2\bigg] \notag\\
  +& \sum_{n \in \{1, 3 \dots\}}^{n_c}
  \bigg[1+\Big(\frac{n\omega_c}{\omega}\Big)^2\bigg] \Bigg\}. \label{m3}
\end{align}
This expression shows that the transfer of angular momentum is governed by the same cavity
parity selection rules in the heat transfer. The odd cavity modes possess electric-field
antinodes in the transverse plane; thus, they drive the torque via the parallel field with
scaling $(1+n^2\omega_c^2/\omega^2)$.
The even modes possess longitudinal electric-field antinodes at the cavity center,
mediating the angular momentum transfer through the perpendicular field with scaling
$(1-n^2\omega_c^2/\omega^2)$.
In the sub-wavelength regime with $L \ll \lambda_{\rm th}$, both Eqs.~\eqref{m2} and
\eqref{m3} reduce to the same asymptotic limit, given by
\begin{equation} \label{m4}
  m_y(\omega) = - \hbar k_0^2 N_d(\omega) \alpha_a'' / L . 
\end{equation}

\begin{figure}
\centering
\includegraphics[width=\columnwidth]{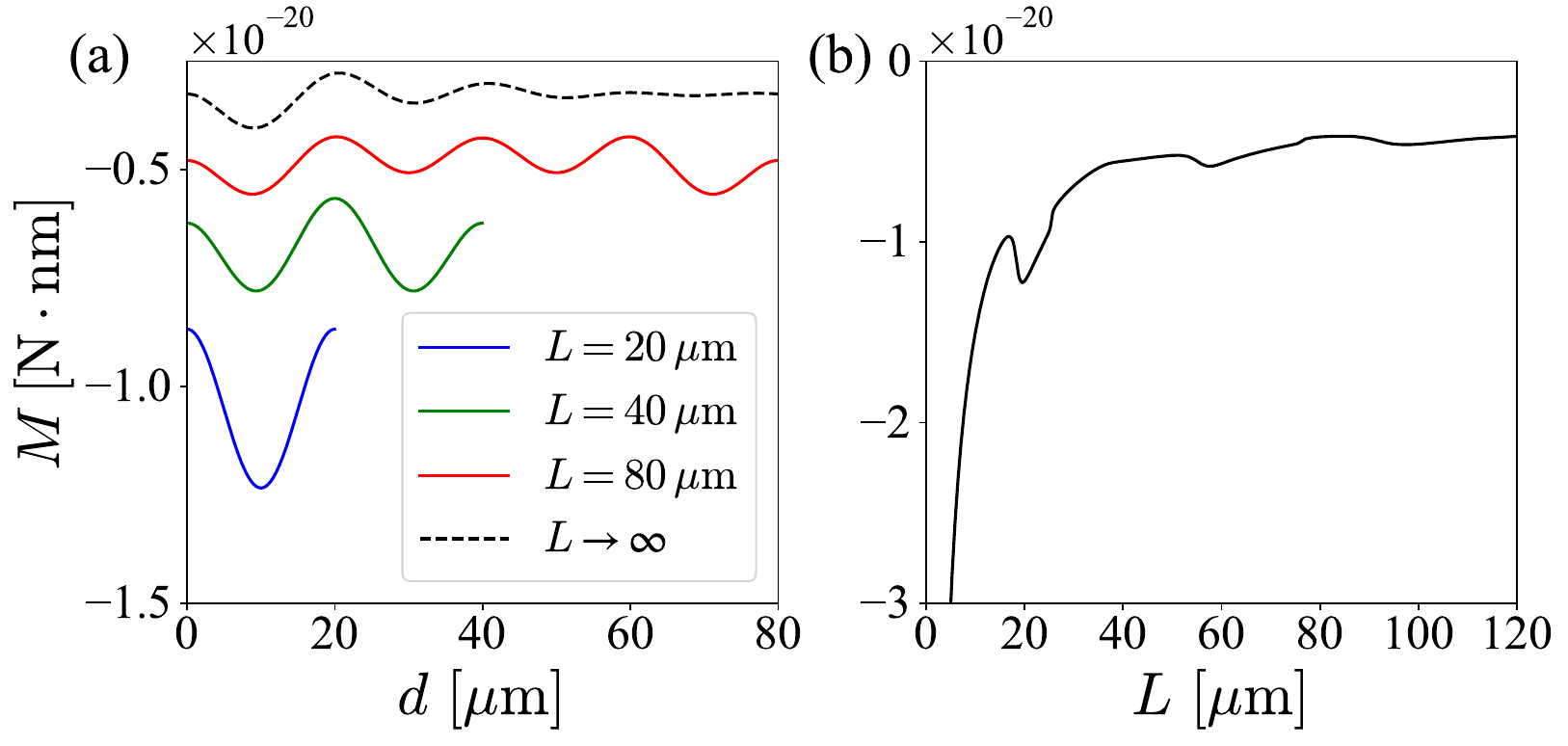} \\
\caption{(a) Torque acting on the nanoparticle as a function of the distance $d$ for
various lengths $L$. (b) Torque at the cavity center as a function of $L$. }
\label{fig4}
\end{figure}

In Fig.~\ref{fig4}(a), we show the torque acting on the magneto-optic nanoparticle as a
function of distance $d$ by varying cavity length $L$.
Unlike the lateral force, the torque remains finite throughout the cavity, since it is
associated with the transfer of spin angular momentum carried by elliptically polarized
thermal photons and can persist even in a spatially symmetric electromagnetic environment.
As in the heat and linear-momentum transfer, the torque exhibits spatial oscillations
induced by the standing-wave structure of the cavity modes. Nevertheless, the torque
preserves the same sign throughout the cavity because all contributing modes transfer
angular momentum with the same handedness determined by the nonreciprocal polarizability
component.

A particular feature is the enhancement of the torque under geometric confinement as shown
in Fig.~\ref{fig4}(b). The torque increases rapidly as the cavity length decreases as
predicted by Eq.~\eqref{m4}. As with the heat transfer, this originates from the
cavity-induced enhancement of the LDOS. In the sub-wavelength regime with $L \ll
\lambda_{\rm th}$, all higher-order cavity modes are beyond the thermally accessible
spectral window, leaving only the zeroth-order mode. Furthermore, as the cavity length
becomes comparable to the thermal wavelength, the torque develops a sequence of
oscillations, which arise from the successive activation of higher-order cavity modes.

\section{Conclusion}\label{summary}
In summary, we studied the exchange of radiative heat, linear momentum, and angular
momentum between a magneto-optic nanoparticle and a Fabry-P\'{e}rot cavity. Within the
framework of fluctuational electrodynamics, we obtained general expressions for the
spectral densities, and evaluated them in several geometric limits. We
showed that geometric confinement affects these transport channels in distinct ways.
Specifically, the thermal Purcell effect enhances radiative heat flow and angular-momentum
transfer, yet it generally suppresses the lateral force. In addition, we demonstrated that
the interference of the cavity modes produces spatial oscillations. 
At the cavity center, spatial parity decouples the field fluctuations into distinct odd
and even contributions. This causes the lateral force to vanish, whereas heat and spin
angular momentum transfer remain finite.
These findings show that cavity-modified electromagnetic fluctuations can be tailored to
manipulate energy and momentum transfers at the nanoscale.

\begin{acknowledgments}
L.J., Y.L., and G.T. are supported by Science Challenge Project (Grant No. TZ2025017) and
National Natural Science Foundation of China (Grants No. 12088101 and No. 12374048).
\end{acknowledgments}

\appendix
\numberwithin{equation}{section}

\section{Magneto-optic nanoparticle} \label{AppendixA}
\renewcommand{\theequation}{A.\arabic{equation}}
We consider the magneto-optic medium InSb where the magnetic field $B$ is applied along
the positive $y$ direction. With the Drude model accounting for both phonon resonances and
free-carrier responses, the frequency-dependent dielectric tensor of the InSb medium takes
the anisotropic form ~\cite{InSb_76, Ott19}
\begin{equation} 
  \epsilon(\omega) =
  \begin{bmatrix}
    \epsilon_d & 0 & i\epsilon_a \\
    0 & \epsilon_y & 0 \\
    -i\epsilon_a & 0 & \epsilon_d
  \end{bmatrix} ,
\end{equation}
where the explicit expressions for these tensor elements are given by
\begin{align*}
  \epsilon_d &= \epsilon_{\infty} \bigg\{ 1+ \frac{\omega_L^2 -\omega_T^2}{\omega_T^2
  -\omega^2 -i\Gamma\omega} + \frac{\omega_{pl}^2(\omega+i\gamma)}{\omega[ \omega_B^2
  -(\omega+i\gamma)^2]} \bigg\}, \\
  \epsilon_y &= \epsilon_{\infty} \bigg[ 1+ \frac{\omega_L^2 -\omega_T^2}{\omega_T^2
  -\omega^2 -i\Gamma\omega} - \frac{\omega_{pl}^2}{\omega(\omega+i\gamma)} \bigg], \\
  \epsilon_a &= \frac{\epsilon_{\infty}\omega_{pl}^2\omega_B}{\omega\big[\omega_B^2
  -(\omega+i\gamma)^2\big]} .
\end{align*}
All the parameters are adopted from Ref.~\cite{InSb_76} with
the high-frequency dielectric constant $\epsilon_\infty = 15.7$, 
longitudinal optical phonon frequency $\omega_L = 3.62\times 10^{13}\,{\rm rad/s}$, 
transverse optical phonon frequency $\omega_T = 3.39\times 10^{13}\,{\rm rad/s}$, 
phonon damping constant $\Gamma = 5.65\times 10^{11}\,{\rm rad/s}$, 
free-carrier damping constant $\gamma = 3.39\times 10^{12}\,{\rm rad/s}$, 
bulk plasma frequency $\omega_{pl} = 3.14\times 10^{13}\,{\rm rad/s}$, and
cyclotron frequency $\omega_B = 8.02\times 10^{12}\,{\rm rad/s}$ corresponding to a 
magnetic field strength $B=1\,$T.

Under the electric-dipole approximation, which is valid when the nanoparticle radius $R$
is much smaller than both the distance to the cavity mirrors and the thermal wavelength
($R \ll d, \hbar c / k_B T$), the polarizability tensor of the spherical InSb nanoparticle
can be derived via the generalized Clausius-Mossotti relation as
\begin{equation}
  \alpha = 4\pi R^3 (\epsilon - \bm{I}) (\epsilon + 2\bm{I})^{-1} =
  \begin{bmatrix}
    \alpha_d & 0 & i\alpha_a \\
    0 & \alpha_y & 0 \\
    -i\alpha_a & 0 & \alpha_d
  \end{bmatrix} ,
\end{equation}
with $\bm{I}$ the $3\times 3$ identity matrix. The components are explicitly expressed as
\begin{align*}
  \alpha_y &= 4\pi R^3 \frac{\epsilon_y - 1}{\epsilon_y + 2} , \\
  \alpha_d &= 4\pi R^3 \left[ 1 - \frac{3(\epsilon_d + 2)}{(\epsilon_d + 2)^2 -
  \epsilon_a^2} \right], \\
  \alpha_a &= 4\pi R^3 \frac{3\epsilon_a}{(\epsilon_d + 2)^2 - \epsilon_a^2} .
\end{align*}
For concise notation, each component is decomposed into its real and imaginary parts as
$\alpha_j = \alpha_j' + i\alpha_j''$ with $j \in \{a,d,y\}$.

\section{Derivation of the spectral densities} \label{AppendixB}
\renewcommand{\theequation}{B.\arabic{equation}}
In this Appendix, we present the derivations of the spectral densities given in
Eqs.~\eqref{h_cavity}, \eqref{f_cavity}, and \eqref{m_cavity}.
Substituting the expressions for the induced dipoles and fields in Eqs.~\eqref{p_in} and
\eqref{E_in}, and the symmetrized cross-correlations in Eqs.~\eqref{EjEk} and
\eqref{pjpk} into Eq.~\eqref{Eqh}, the spectral density of the net radiated power from the
nanoparticle to the environment evaluates to
\begin{align}
  h(\omega) &= 2\sum_k \int_{\omega'} \omega {\rm Im}\big[\big\langle 
  p_k^{\rm fl^*}(\omega)E_k^{\rm in}(\omega') \big\rangle 
  +({\rm in} \leftrightarrow {\rm fl}) \big] \notag \\
  &= 2\sum_{kl}\int_{\omega'}\omega{\rm Im}\big[\mu_0\omega'^2\big\langle p_k^{\rm
      fl^*}(\omega)
  p_l^{\rm fl}(\omega')\big\rangle G_{kl}(\bm{r},\bm{r},\omega') \notag\\
  &\qquad\qquad\qquad\ +\epsilon_0\alpha_{kl}^*\big\langle E_k^{\rm fl}(
  \omega') E_l^{\rm fl^*}(\omega) \big\rangle\big] \notag \\
  &= 4\hbar \omega k_0^2N_d(\omega)\big\{ \alpha_d''{\rm
  Im}[G_{xx}(\bm{r},\bm{r})+G_{zz}(\bm{r},\bm{r})]\notag\\
  &\qquad\qquad\qquad +\alpha_y''{\rm Im}[G_{yy}(\bm{r},\bm{r})]\big\}.
\end{align}
It is important to emphasize that in the derivation above, the nanoparticle is treated
within the point-dipole approximation. Because the particle size is assumed to be much
smaller than both the thermal wavelength and the distances to the cavity mirrors, the
spatial variations of the field across its volume are negligible. Consequently, the dyadic
Green's function is evaluated at the spatial coordinate of the nanoparticle.
The required Green's function in real space can be obtained by inserting
Eq.~\eqref{G_cavity} into the Fourier transform in Eq.~\eqref{G_FT}.

Similarly, using Eq.~\eqref{Eqf} and expressing the spatial derivative $\partial_x$ in
terms of the in-plane wavevector $q_x$, the spectral density of the lateral force acting
on the nanoparticle is derived as
\begin{align}
  & f_x(\omega) = 2\sum_k\int_{\omega'} {\rm Re}\big[\big\langle 
  p_k^{\rm fl}(\omega) \partial_x E_k^{\rm in^*}(\omega')\big\rangle
  +({\rm in} \leftrightarrow {\rm fl}) \big] \notag \\
  =& 2\sum_{kl}\int_{\omega'} {\rm Re}\big\{ \mu_0\omega'^2 \langle p_k^{\rm fl}(\omega)
    p_l^{\rm fl^*}(\omega')\rangle \partial_x G_{kl}^*(\bm{r},\bm{r},\omega') \notag \\
  & \qquad\qquad\quad + \epsilon_0\alpha_{kl}(\omega)\langle E_l^{\rm fl}(\omega)
  \partial_x E_k^{\rm fl^*}(\omega')\rangle \big\} \notag \\
  =& 4\hbar k_0^2 N_d(\omega)\alpha_a'' {\int \frac{d^2\bm{q}}{(2\pi)^2}} q_x {\rm Re}
  [G_{xz}(\bm{q}, z, z) -G_{zx}(\bm{q}, z, z)] .
\end{align}
This expression explicitly shows that the lateral force is mediated by the off-diagonal
polarizability component $\alpha_a''$, and arises from the nonreciprocal cross-coupling
between the electromagnetic modes along the $x$- and $z$-directions.

Finally, the spectral density of the net torque acting on the nanoparticle yields
\begin{align}
  m_y(\omega) &= 2\sum_{kl}\int_{\omega'} \epsilon_{ykl} {\rm Re}\big[\big< p_k^{\rm
  fl}(\omega) E_l^{\rm in^*}(\omega')\big> +({\rm in}\leftrightarrow {\rm fl}) \big]
  \notag \\
  =& 2\sum_{jkl}\int_{\omega'}\epsilon_{ykl}{\rm Re}\big[\mu_0\omega'^2\big<
  p_k^{\rm fl}(\omega) p_j^{\rm fl^*}(\omega')\big> G_{lj}^*(\bm{r},\bm{r},\omega')
  \notag \\
  &\qquad\qquad\qquad\quad + \epsilon_0\alpha_{kj}\big\langle
  E_j^{\rm fl}(\omega)E_l^{\rm fl^*}(\omega')\big\rangle \big] \notag \\
  =& -4\hbar k_0^2 N_d(\omega)\alpha_a''{\rm Im}
  \big[G_{xx}(\bm{r},\bm{r})+G_{zz}(\bm{r},\bm{r})\big].
\end{align}
Unlike the lateral force, which requires asymmetric mode coupling, the torque depends only
on the diagonal components of the Green's function. Therefore, angular momentum transfer
can occur in spatially symmetric photonic environments.

\bibliography{bib_heat_radiation}{}

@article{Sipe87,
  author = {J. E. Sipe},
  journal = {J. Opt. Soc. Am. B},
  number = {4},
  pages = {481--489},
  publisher = {Optica Publishing Group},
  title = {New {G}reen-function formalism for surface optics},
  volume = {4},
  month = {Apr},
  year = {1987},
  url = {https://opg.optica.org/josab/abstract.cfm?URI=josab-4-4-481},
  doi = {10.1364/JOSAB.4.000481}
}

@article{Joulain03,
  title = {Definition and measurement of the local density of electromagnetic states close
           to an interface},
  author = {Joulain, Karl and Carminati, R\'emi and Mulet, Jean-Philippe and Greffet,
            Jean-Jacques},
  journal = {Phys. Rev. B},
  volume = {68},
  issue = {24},
  pages = {245405},
  numpages = {10},
  year = {2003},
  month = {Dec},
  publisher = {American Physical Society},
  doi = {10.1103/PhysRevB.68.245405},
  url = {https://link.aps.org/doi/10.1103/PhysRevB.68.245405}
}

@article{Pendry99,
	author = {J B Pendry},
	title = {Radiative exchange of heat between nanostructures},
  journal = {J. Phys.: Condens. Matter},
	volume = {11},
	number = {35},
	pages = {6621--6633},
	year = 1999,
	month = {aug},
	publisher = {{IOP} Publishing},
	doi = {10.1088/0953-8984/11/35/301},
	url = {https://doi.org/10.1088/0953-8984/11/35/301},
}

@article{review05,
  title={Surface electromagnetic waves thermally excited: Radiative heat transfer,
         coherence properties and {C}asimir forces revisited in the near field},
  author={Karl Joulain and Jean-Philippe Mulet and Franccois Marquier and R'emi Carminati
          and Jean-Jacques Greffet},
  journal = {Surf. Sci. Rep.},
  volume = {57},
  number = {3},
  pages = {59-112},
  year={2005},
  doi = {https://doi.org/10.1016/j.surfrep.2004.12.002},
  url = {http://www.sciencedirect.com/science/article/pii/S0167572905000105},
}

@article{review07,
  title = {Near-field radiative heat transfer and noncontact friction},
  author = {Volokitin, A. I. and Persson, B. N. J.},
  journal = {Rev. Mod. Phys.},
  volume = {79},
  issue = {4},
  pages = {1291--1329},
  numpages = {0},
  year = {2007},
  month = {Oct},
  publisher = {American Physical Society},
  doi = {10.1103/RevModPhys.79.1291},
  url = {https://link.aps.org/doi/10.1103/RevModPhys.79.1291}
}

@article{review15,
  author = {Song,Bai  and Fiorino,Anthony and Meyhofer,Edgar and Reddy,Pramod},
  title = {Near-field radiative thermal transport: From theory to experiment},
  journal = {AIP Adv.},
  volume = {5},
  number = {5},
  pages = {053503},
  year = {2015},
  doi = {10.1063/1.4919048},
  URL = {https://doi.org/10.1063/1.4919048},
}

@article{review18,
  author = {Cuevas, Juan Carlos and García-Vidal, Francisco J.},
  title = {Radiative Heat Transfer},
  journal = {ACS Photonics},
  volume = {5},
  number = {10},
  pages = {3896-3915},
  year = {2018},
  doi = {10.1021/acsphotonics.8b01031},
  URL = {https://doi.org/10.1021/acsphotonics.8b01031},
}

@article{review21,
  title = {Near-field radiative heat transfer in many-body systems},
  author = {Biehs, S.-A. and Messina, R. and Venkataram, P. S. and Rodriguez, A. W. and
            Cuevas, J. C. and Ben-Abdallah, P.},
  journal = {Rev. Mod. Phys.},
  volume = {93},
  issue = {2},
  pages = {025009},
  numpages = {51},
  year = {2021},
  month = {Jun},
  publisher = {American Physical Society},
  doi = {10.1103/RevModPhys.93.025009},
  url = {https://link.aps.org/doi/10.1103/RevModPhys.93.025009}
}

@article{nonrecip_rotate19,
  title = {Magnetically activated rotational vacuum friction},
  author = {Pan, Deng and Xu, Hongxing and Garc\'{\i}a de Abajo, F. Javier},
  journal = {Phys. Rev. A},
  volume = {99},
  issue = {6},
  pages = {062509},
  numpages = {8},
  year = {2019},
  month = {Jun},
  publisher = {American Physical Society},
  doi = {10.1103/PhysRevA.99.062509},
  url = {https://link.aps.org/doi/10.1103/PhysRevA.99.062509}
}

@article{lateral_17,
  title = {Fluctuation-induced forces in the presence of mobile carrier drift},
  author = {Shapiro, Boris},
  journal = {Phys. Rev. B},
  volume = {96},
  issue = {7},
  pages = {075407},
  numpages = {12},
  year = {2017},
  month = {Aug},
  publisher = {American Physical Society},
  doi = {10.1103/PhysRevB.96.075407},
  url = {https://link.aps.org/doi/10.1103/PhysRevB.96.075407}
}

@article{lateral_21,
  title = {Near Field Propulsion Forces from Nonreciprocal Media},
  author = {Gelbwaser-Klimovsky, David and Graham, Noah and Kardar, Mehran and Kr\"uger,
            Matthias},
  journal = {Phys. Rev. Lett.},
  volume = {126},
  issue = {17},
  pages = {170401},
  numpages = {5},
  year = {2021},
  month = {Apr},
  publisher = {American Physical Society},
  doi = {10.1103/PhysRevLett.126.170401},
  url = {https://link.aps.org/doi/10.1103/PhysRevLett.126.170401}
}

@article{lateral_21-2,
  title = {Nonequilibrium lateral force and torque by thermally excited nonreciprocal
           surface electromagnetic waves},
  author = {Khandekar, Chinmay and Buddhiraju, Siddharth and Wilkinson, Paul R. and
            Gimzewski, James K. and Rodriguez, Alejandro W. and Chase, Charles and Fan,
            Shanhui},
  journal = {Phys. Rev. B},
  volume = {104},
  issue = {24},
  pages = {245433},
  numpages = {11},
  year = {2021},
  month = {Dec},
  publisher = {American Physical Society},
  doi = {10.1103/PhysRevB.104.245433},
  url = {https://link.aps.org/doi/10.1103/PhysRevB.104.245433}
}

@article{lateral_23,
  title = {Moving media as photonic heat engine and pump},
  author = {Tsurimaki, Yoichiro and Yu, Renwen and Fan, Shanhui},
  journal = {Phys. Rev. B},
  volume = {107},
  issue = {11},
  pages = {115406},
  numpages = {15},
  year = {2023},
  month = {Mar},
  publisher = {American Physical Society},
  doi = {10.1103/PhysRevB.107.115406},
  url = {https://link.aps.org/doi/10.1103/PhysRevB.107.115406}
}

@article{lateral_23-2,
  title = {Vacuum torque, propulsive forces, and anomalous tangential forces: Effects of nonreciprocal media out of thermal equilibrium},
  author = {Milton, Kimball A. and Guo, Xin and Kennedy, Gerard and Pourtolami, Nima and DelCol, Dylan M.},
  journal = {Phys. Rev. A},
  volume = {108},
  issue = {2},
  pages = {022809},
  numpages = {15},
  year = {2023},
  month = {Aug},
  publisher = {American Physical Society},
  doi = {10.1103/PhysRevA.108.022809},
  url = {https://link.aps.org/doi/10.1103/PhysRevA.108.022809}
}

@article{GT24-2,
  title = {Current-induced near-field radiative energy, linear-momentum, and
           angular-momentum transfer},
  author = {Zhu, Huimin and Tang, Gaomin and Zhang, Lei and Chen, Jun},
  journal = {Phys. Rev. B},
  volume = {109},
  issue = {7},
  pages = {075413},
  numpages = {7},
  year = {2024},
  month = {Feb},
  publisher = {American Physical Society},
  doi = {10.1103/PhysRevB.109.075413},
  url = {https://link.aps.org/doi/10.1103/PhysRevB.109.075413}
}

@article{lateral_25,
  title = {Propulsion force and heat transfer for nonreciprocal nanoparticles},
  author = {Henkes, Laila and Asheichyk, Kiryl and Kr\"uger, Matthias},
  journal = {Phys. Rev. B},
  volume = {111},
  issue = {3},
  pages = {035441},
  numpages = {13},
  year = {2025},
  month = {Jan},
  publisher = {American Physical Society},
  doi = {10.1103/PhysRevB.111.035441},
  url = {https://link.aps.org/doi/10.1103/PhysRevB.111.035441}
}

@article{GT26-1,
  title = {Modulating near-field radiative energy and momentum transfer via rotating
           {W}eyl semimetals},
  author = {Zhu, Huimin and Tang, Gaomin and Zhang, Lei and Chen, Jun},
  journal = {Phys. Rev. B},
  volume = {113},
  issue = {12},
  pages = {L121404},
  numpages = {6},
  year = {2026},
  month = {Mar},
  publisher = {American Physical Society},
  doi = {10.1103/15hd-v4kp},
  url = {https://link.aps.org/doi/10.1103/15hd-v4kp}
}

@article{InSb_76,
  title = {Coupled surface magnetoplasmon-optic-phonon polariton modes on {I}n{S}b},
  author = {Palik, E. D. and Kaplan, R. and Gammon, R. W. and Kaplan, H. and Wallis, R. F.
            and Quinn, J. J.},
  journal = {Phys. Rev. B},
  volume = {13},
  issue = {6},
  pages = {2497--2506},
  numpages = {0},
  year = {1976},
  month = {Mar},
  publisher = {American Physical Society},
  doi = {10.1103/PhysRevB.13.2497},
  url = {https://link.aps.org/doi/10.1103/PhysRevB.13.2497}
}

@article{WSM_radiate1,
  author = {Zhao, Bo and Guo, Cheng and Garcia, Christina A. C. and Narang, Prineha and
            Fan, Shanhui},
  title = {Axion-Field-Enabled Nonreciprocal Thermal Radiation in {W}eyl Semimetals},
  journal = {Nano Lett.},
  volume = {20},
  number = {3},
  pages = {1923-1927},
  year = {2020},
  doi = {10.1021/acs.nanolett.9b05179},
  URL = {https://doi.org/10.1021/acs.nanolett.9b05179},
}

@article{WSM_radiate2,
  author = {Guo, Cheng and Zhao, Bo and Huang, Danhong and Fan, Shanhui},
  title = {Radiative Thermal Router Based on Tunable Magnetic {W}eyl Semimetals},
  journal = {ACS Photonics},
  volume = {7},
  number = {11},
  pages = {3257-3263},
  year = {2020},
  doi = {10.1021/acsphotonics.0c01376},
  URL = {https://doi.org/10.1021/acsphotonics.0c01376},
}

@article{WSM_radiate3,
  title = {Large nonreciprocal absorption and emission of radiation in type-{I} {W}eyl
           semimetals with time reversal symmetry breaking},
  author = {Tsurimaki, Yoichiro and Qian, Xin and Pajovic, Simo and Han, Fei and Li,
            Mingda and Chen, Gang},
  journal = {Phys. Rev. B},
  volume = {101},
  issue = {16},
  pages = {165426},
  numpages = {11},
  year = {2020},
  month = {Apr},
  publisher = {American Physical Society},
  doi = {10.1103/PhysRevB.101.165426},
  url = {https://link.aps.org/doi/10.1103/PhysRevB.101.165426}
}

@article{WSM_radiate4,
  title = {Intrinsic nonreciprocal reflection and violation of {K}irchhoff's law of
           radiation in planar type-{I} magnetic {W}eyl semimetal surfaces},
  author = {Pajovic, Simo and Tsurimaki, Yoichiro and Qian, Xin and Chen, Gang},
  journal = {Phys. Rev. B},
  volume = {102},
  issue = {16},
  pages = {165417},
  numpages = {11},
  year = {2020},
  month = {Oct},
  publisher = {American Physical Society},
  doi = {10.1103/PhysRevB.102.165417},
  url = {https://link.aps.org/doi/10.1103/PhysRevB.102.165417}
}

@article{GT_WSM,
  author = {Tang, Gaomin and Chen, Jun and Zhang, Lei},
  title = {Twist-induced control of near-field heat radiation between magnetic {W}eyl
           semimetals},
  journal = {ACS Photonics},
  volume = {8},
  number = {2},
  pages = {443-448},
  year = {2021},
  doi = {10.1021/acsphotonics.0c01945},
  URL = {https://doi.org/10.1021/acsphotonics.0c01945},
}

@article{cavity-23,
  author={Jarc, Giacomo and Mathengattil, Shahla Yasmin and Montanaro, Angela and Giusti,
          Francesca and Rigoni, Enrico Maria and Sergo, Rudi and Fassioli, Francesca and
          Winnerl, Stephan and Dal Zilio, Simone and Mihailovic, Dragan and
          Prelov{\v{s}}ek, Peter and Eckstein, Martin and Fausti, Daniele},
  title={Cavity-mediated thermal control of metal-to-insulator transition in
         1{T}-{T}a{S}$_2$},
  journal={Nature},
  year={2023},
  month={Oct},
  day={01},
  volume={622},
  number={7983},
  pages={487-492},
  issn={1476-4687},
  doi={10.1038/s41586-023-06596-2},
  url={https://doi.org/10.1038/s41586-023-06596-2}
}

@article{cavity-24,
  title = {Thermal {P}urcell effect and cavity-induced renormalization of dissipations},
  author = {Chiriac\`o, Giuliano},
  journal = {Phys. Rev. B},
  volume = {110},
  issue = {16},
  pages = {L161107},
  numpages = {7},
  year = {2024},
  month = {Oct},
  publisher = {American Physical Society},
  doi = {10.1103/PhysRevB.110.L161107},
  url = {https://link.aps.org/doi/10.1103/PhysRevB.110.L161107}
}

@article{cavity-25,
  title = {Controlling radiative heat flow through cavity electrodynamics},
  author = {Fassioli, Francesca and Faist, Jerome and Eckstein, Martin and Fausti,
            Daniele},
  journal = {Phys. Rev. B},
  volume = {111},
  issue = {16},
  pages = {165425},
  numpages = {6},
  year = {2025},
  month = {Apr},
  publisher = {American Physical Society},
  doi = {10.1103/PhysRevB.111.165425},
  url = {https://link.aps.org/doi/10.1103/PhysRevB.111.165425}
}

@article{cavity-25-2,
  title = {Nonthermal electron-photon steady states in open cavity quantum materials},
  author = {Flores-Calder\'on, R. and Islam, Md Mursalin and Pini, Michele and Piazza,
            Francesco},
  journal = {Phys. Rev. Res.},
  volume = {7},
  issue = {1},
  pages = {013073},
  numpages = {6},
  year = {2025},
  month = {Jan},
  publisher = {American Physical Society},
  doi = {10.1103/PhysRevResearch.7.013073},
  url = {https://link.aps.org/doi/10.1103/PhysRevResearch.7.013073}
}

@article{GT26-2,
  title = {Transverse Thermophotovoltaics from Nonreciprocal Plasmon Drag in Metal},
  author = {He, Dingwei and Tang, Gaomin},
  journal = {Phys. Rev. Lett.},
  volume = {136},
  issue = {17},
  pages = {176901},
  numpages = {8},
  year = {2026},
  month = {Apr},
  publisher = {American Physical Society},
  doi = {10.1103/kl16-bw43},
  url = {https://link.aps.org/doi/10.1103/kl16-bw43}
}

@article{Ott19,
  author = {Ott, Annika and Messina, Riccardo and Ben-Abdallah, Philippe and Biehs,
            Svend-Age},
  title = {Radiative thermal diode driven by nonreciprocal surface waves},
  journal = {Appl. Phys. Lett.},
  volume = {114},
  number = {16},
  pages = {163105},
  year = {2019},
  month = {04},
  issn = {0003-6951},
  doi = {10.1063/1.5093626},
  url = {https://doi.org/10.1063/1.5093626}
}

\end{document}